# Superconductivity in PtSbS with a Noncentrosymmetric Cubic Crystal Structure

Ryosuke Mizutani[1], Yoshihiko Okamoto[1,*], Hayate Nagaso[1], Youichi Yamakawa[2], Hiroshi Takatsu[3], Hiroshi Kageyama[3], Shunichiro Kittaka[4], Yohei Kono[4], Toshiro Sakakibara[4], and Koshi Takenaka[1]

[1]*Department of Applied Physics, Nagoya University, Nagoya 464-8603, Japan*
[2]*Department of Physics, Nagoya University, Nagoya 464-8602, Japan*
[3]*Graduate School of Engineering, Kyoto University, Kyoto 615-8510, Japan*
[4]*Institute for Solid State Physics, University of Tokyo, Kashiwa 277-8581, Japan*

We report the synthesis, electronic properties, and electronic structure of ullmannite-type PtSbS, which has a cubic crystal structure without space inversion symmetry. Electrical resistivity and magnetization measured at low temperatures suggested that this compound is a bulk superconductor with a superconducting transition temperature of $T_c$ = 0.15 K. First principles calculations indicated that Fermi surfaces of PtSbS include strongly nested hole pockets, which can make this compound interesting if they contribute to the emergence of superconductivity.

Platinum is a *d*-block element with the third largest atomic number, after mercury and gold, among the experimentally available elements. Platinum-based superconductors have been intensively studied in recent years as a platform where unconventional superconductors can be realized due to the strong spin–orbit coupling coming from the large atomic number and the orbital degree of freedom of the 5*d* electrons. The most typical example is Li$_2$Pt$_3$B, which has a noncentrosymmetric crystal structure with the cubic $P4_332$ space group. It becomes superconducting below $T_c$ = 2 K.[1] Magnetic penetration depth, $^{11}$B- and $^{195}$Pt-NMR, and heat capacity data suggested that the spin-triplet Cooper pairs with a line node are dominant in the superconducting phase.[2-4] On the other hand, half-Heusler-type YPtBi with $T_c$ = 0.77 K[5] has recently attracted because of the pairing of $J$ = 3/2 fermions. In YPtBi, the strong spin–orbit coupling causes band inversion, which results in the $J$ = 3/2 state being dominant at around the Fermi energy[6,7] and responsible for various types of topological superconductivity.[8-10] Common to both materials, the crystal structures are cubic and do not have space inversion symmetry. Strong spin–orbit coupling acting on a noncentrosymmetric system and the multiorbital nature of 5*d* electrons are expected to play an important role in the realization of unconventional superconductivity. Additionally, unconventional Cooper pairing, such as chiral $d+id$ wave, was discussed in hexagonal SrPtAs,[11-13] while tetragonal SrPt$_3$P was suggested to be a strong-coupling multiband superconductor.[14,15] It is thus clear that exploring platinum compounds with high crystal symmetry is promising for discovering unconventional superconductivity.

In this letter, we report that noncentrosymmetric cubic PtSbS is a candidate of unconventional superconductor. PtSbS crystallizes in the ullmannite-type structure with the cubic $P2_13$ space group, as shown in Fig. 1(a).[16] This crystal structure is an ordered pyrite type, where the anion sites in pyrite-type MX$_2$ are occupied by two kind of atoms. Many compounds with the group 9 and 10 transition metal elements have this crystal structure.[16,17] It is reported that the compounds with group 9 elements are semiconducting, while those with group 10 elements, including PtSbS, are metallic.[17] Two of the latter compounds, PdBiSe and PtBiSe, showed a superconducting transition at ~1 K.[18] However, results denying bulk superconductivity were later shown for PdBiSe[19] and bulk superconductivity has not been confirmed for PtBiSe. Here, PtSbS is strongly suggested to be a bulk superconductor with a transition temperature of $T_c = 0.15$ K based on the resistivity and magnetization measurements at low temperatures. Zero resistivity was observed below 0.15 K by electrical resistivity measurements using adiabatic demagnetization cooling, and large diamagnetic magnetization was observed below 0.12 K by DC magnetization measurements using a dilution refrigerator. First principles calculations have revealed that PtSbS has Fermi surfaces with characteristic shapes being different from those of isostructural NiSbS and PdBiSe, indicating that the electronic structures of the ullmannite-type compounds are sensitive to the period of the constituent elements and chemical pressure. Therefore, ullmannite-type compounds are promising as a platform to realize various interesting electronic properties, such as unconventional superconductivity.

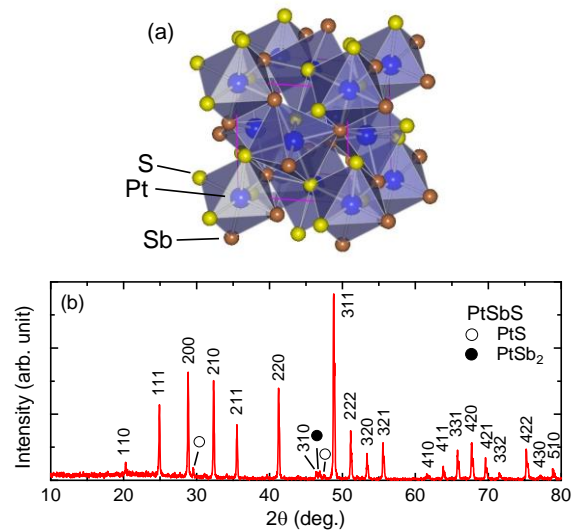

Fig. 1. (a) Crystal structure of PtSbS.[20] The solid line indicates the unit cell. (b) Powder XRD pattern of a PtSbS polycrystalline sample taken at room temperature, indexed by a cubic cell of $a$ = 6.174 Å. The peaks indicated by open and filled circles are those from PtS and PtSb$_2$ impurities, respectively.

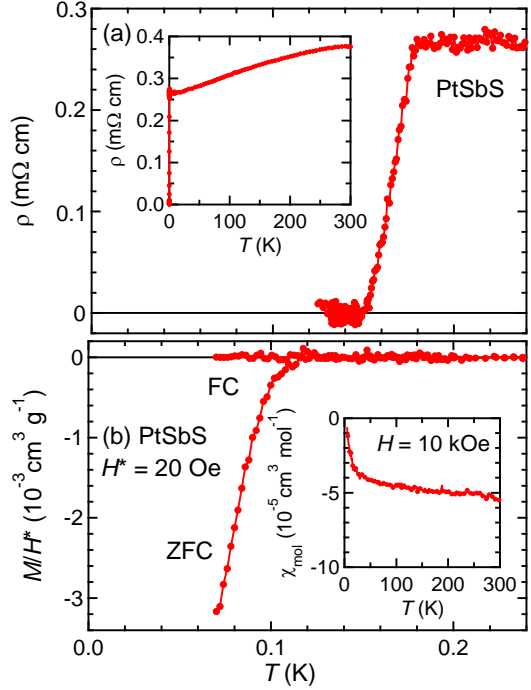

Fig. 2. (a) Temperature dependence of electrical resistivity of the PtSbS polycrystalline sample. The inset shows the data up to room temperature. (b) Temperature dependences of field-cooled and zero-field-cooled magnetizations measured at the effective magnetic field of $H^* = 20$ Oe. The inset shows the temperature dependence of magnetic susceptibility measured at a magnetic field of 10 kOe.

Polycrystalline samples of PtSbS were prepared by a solid-state reaction method. An equimolar mixture of Pt (99.9%, RARE METALLIC), Sb (99.9%, Kojundo Chemical Lab.), and S (99.99%, Kojundo Chemical Lab.) powders was pressed and sealed in an evacuated quartz tube. The tube was first kept at 400 °C for 24 h to prevent rapid vaporization of sulfur, at 750 °C for 3 h, and then at 500 °C for 24 h. Thereafter, the tube was furnace-cooled to room temperature. Sample characterization was performed by powder X-ray diffraction (XRD) analysis with Cu Kα radiation at room temperature using a RINT-2100 diffractometer (RIGAKU). As shown in Fig. 1(b), all diffraction peaks, except some small peaks due to PtS and $PtSb_2$ impurities, were indexed on the basis of a cubic cell of $a = 6.174$ Å, indicating that the ullmannite-type PtSbS was obtained as the main phase.[16]

Electrical resistivity measurements down to 0.1 K were performed using an adiabatic demagnetization refrigerator. DC magnetization measurements down to 0.07 K were performed using a Faraday force magnetometer with a dilution refrigerator.[21] Magnetization was measured under a magnetic-field gradient of 1 T m$^{-1}$ together with a centered magnetic field generated by a 15-T magnet. Because the diamagnetic signal of the sample was smallest at an external magnetic field $H$ of 40 Oe, we assumed that the effective magnetic field $H^*$ in the sample becomes zero at this field (i.e., $H^* = H - 40$ Oe). Electrical resistivity, heat capacity, and magnetic susceptibility measurements down to 2 K were performed using PPMS and MPMS (both Quantum Design). First principles calculations were performed using the WIEN2k code.[22] The experimentally obtained structural parameters were used for the calculations.

Figures 2(a) and (b) show the temperature dependences of electrical resistivity $\rho$ and magnetization $M$ of the PtSbS polycrystalline samples, respectively. As seen from the inset of Fig. 2(a), the electrical resistivity data show metallic behavior below room temperature. The $\rho$ decreases with decreasing temperature, shows a sharp drop below 0.18 K, and becomes zero at 0.15 K. Parallel to the drastic decrease of $\rho$, a strong diamagnetic signal with a decrease of $M$ below 0.12 K was observed in the zero-field-cooled magnetization data measured under the effective magnetic field $H^*$ of 20 Oe, as shown in Fig. 2(b), indicating that a superconducting transition occurs at this temperature. Although the onset temperature of the decrease in $M$ is lower than that of $\rho$, this temperature difference is probably due to a residual magnetic field. Providing that the sample is a single phase of PtSbS, a shielding fraction at $T = 70$ mK is estimated to be 39%. Considering that $M$ is not saturated at this temperature and the volume fraction of PtS and $PtSb_2$ impurity phases is up to several percent, as seen in Fig. 1(b), the observed diamagnetic signal strongly suggests that a bulk superconducting transition occurs in PtSbS. In addition, PtS and $PtSb_2$ were reported not to be metallic.[17,23,24] Field-cooled data do not show a significant Meisner signal, which is reasonable considering that PtSbS is a type-II superconductor with pinning. From the zero-resistivity temperature, the superconducting transition temperature $T_c$ of PtSbS is determined to be 0.15 K.

We now discuss the superconducting properties of PtSbS under a magnetic field. Figure 3 shows the temperature dependences of $\rho$ measured at various magnetic fields. The superconducting transition temperature is decreased with increasing the magnetic field and the zero-resistivity $T_c$ is suppressed below 0.1 K by applying a magnetic field of 300 Oe. The upper critical field, $H_{c2}$, of PtSbS obtained from the $\rho$ data is shown in Fig. 3(b). The $H_{c2}$ data are fitted to the Ginzburg-Landau (GL) formula $H_{c2}(T) = H_{c2}(0)[1 - (T/T_c)^2]/[1 + (T/T_c)^2]$, yielding $H_{c2}(0)$ of $6.2(2) \times 10^2$ Oe and a GL coherence length of $\xi_{GL} = 73$ nm. This $H_{c2}(0)$ does not exceed the Pauli-limiting field,[25] the same as in various Pt-based superconductors, with the exception of the heavy-fermion superconductor CePt$_3$Si.

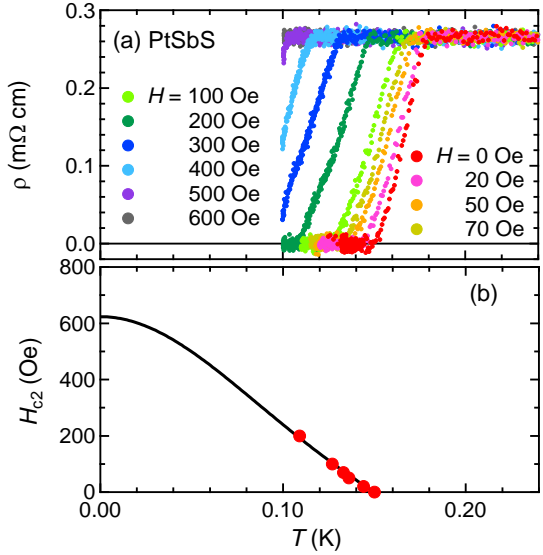

Fig. 3. (a) Temperature dependences of electrical resistivity of a PtSbS polycrystalline sample measured at various magnetic fields of 0–600 Oe. (b) Temperature dependence of the upper critical field $H_{c2}$ determined by the zero-resistivity temperature shown in (a). The solid curve shows a fit to the GL formula.

The electronic structure of PtSbS, which is thus suggested to be a bulk superconductor, is discussed with the experimental data of the normal state. The electronic band structure and electronic density of states (DOS) of PtSbS calculated with spin-orbit coupling are shown in Fig 4(a). As in the cases of NiSbS and PdBiSe,[19,26,27] which are isostructural and isoelectronic to PtSbS, the band structure of PtSbS is metallic because the electronic bands cross the Fermi energy $E_F$. This is consistent with the observed superconducting ground state. As seen in the electronic DOS shown in the right panel of Fig. 4(a), the contribution of Pt at $E_F$ is considerably larger than those of Sb and S. In the region of energy below $E_F$, a band gap of ~1 eV opens below the energy lower than $E_F$ by 0.6 eV. This is probably due to the crystal-field splitting that occurs between Pt 5$d$ bands consisting of $t_{2g}$ and $e_g$ orbitals, reflecting that a Pt atom is octahedrally coordinated, as shown in Fig. 1(a). The formal valence of each atom in PtSbS is Pt$^{3+}$, Sb$^-$, and S$^{2-}$, resulting in the low spin state of the 5$d^7$ electron configuration of Pt. In IrSbS with the same crystal structure and one less electron than in PtSbS, the $E_F$ is likely located in the band gap, because Ir$^{3+}$ (5$d^6$) has the low spin state, as found in IrSbS with semi-conducting properties.[17] These discussions suggest that the above assignment of the valence is appropriate in these ullmannite-type compounds and the electronic states near the $E_F$ in PtSbS mainly consist of the $e_g$-like $5d$ orbitals of Pt atoms.

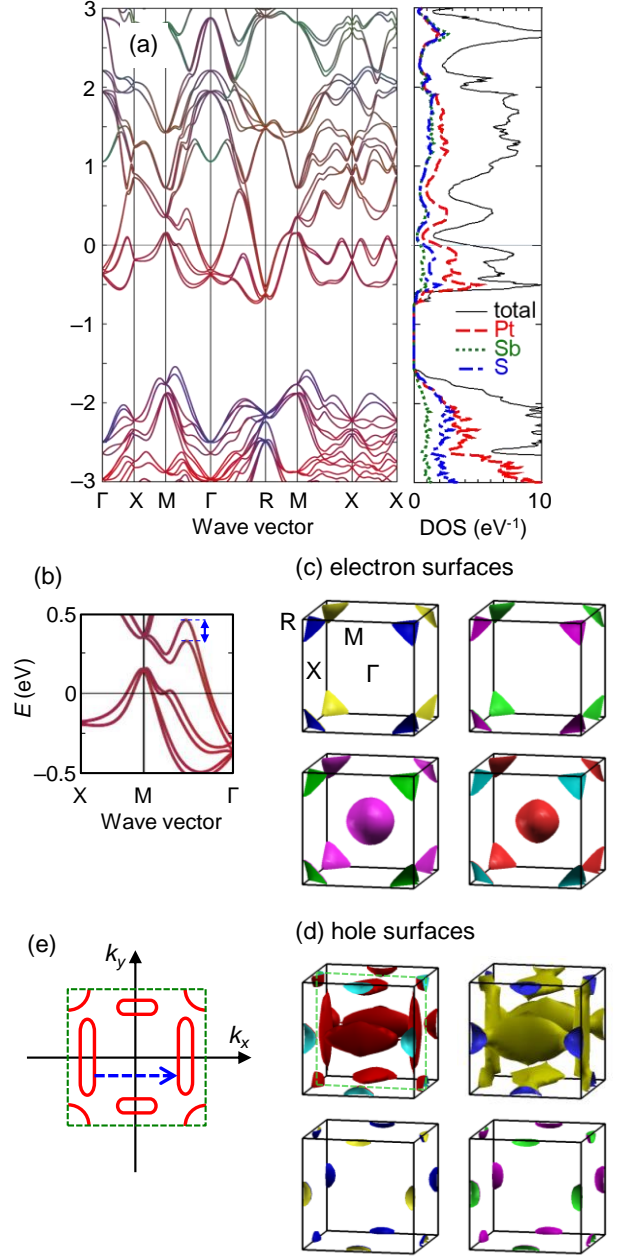

Fig. 4. Electronic structure of PtSbS calculated with spin-orbit coupling. (a) Electronic band structure (left) and total and partial electronic DOS (right). The Fermi level is set to 0 eV. (b) An enlarged view of (a). The largest spin splitting of energy bands in this panel is shown by an arrow. (c, d) Fermi surfaces including four electron surfaces (c) and four hole ones (d). The broken line in the top left panel of (d) indicates the cross section shown in (e). (e) A schematic picture of the cross section at $k_z = 0$ of the hole Fermi surface shown in the top left panel of (d). The broken arrow represents the nesting vector discussed in the main text.

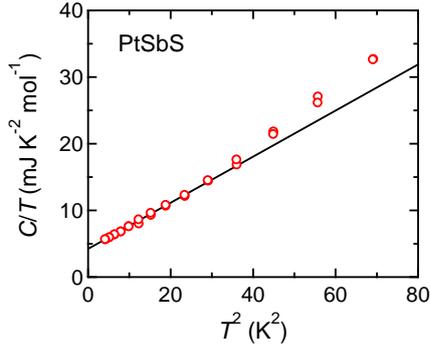

Fig. 5. Heat capacity divided by temperature of a PtSbS polycrystalline sample as a function of $T^2$. The solid line shows a result of the linear fit between 2 and 4.5 K.

The electronic specific heat of PtSbS is calculated to be $\gamma_{band}$ = 2.6 mJ K$^{-2}$ mol$^{-1}$ from the DOS at $E_F$ (4.5 eV$^{-1}$) shown in Fig. 4(a). The heat capacity data shown in Fig. 5 yielded β = 0.346(13) mJ K$^{-4}$ mol$^{-1}$ and an electronic specific heat of γ = 4.22(14) mJ K$^{-2}$ mol$^{-1}$ by a linear fit to $C/T = \beta T^2 + \gamma$ between 2 and 4.5 K. This β value yields Debye temperature of $\theta_D$ = 256 K, which is between 186 K in PdBiSe and 310 K in NiSbS.[19] This result is consistent with the fact that PdBiSe and NiSbS, respectively, have larger and smaller formula weight than that of PtSbS. The obtained γ is ~60% larger than $\gamma_{band}$, suggesting that γ is enhanced by the electron-electron and/or electron-phonon interaction. Usually, one can judge which interaction is dominant for the enhancement of γ by the Wilson ratio, but it cannot be used in the PtSbS case. As shown in the inset of Fig. 2(b), the magnetic susceptibility of the normal state of PtSbS is negative due to the larger diamagnetic contribution from the inner shell electrons than the Pauli paramagnetic one. It is difficult to quantitatively evaluate the value of Pauli paramagnetic susceptibility.

As can be seen from Figs. 4(a) and (b), spin splitting of the energy bands at around $E_F$, which is caused by the antisymmetric spin-orbit coupling, is small in PtSbS, although the electronic state near $E_F$ mainly consist of Pt $5d$ orbitals. The splitting is up to 0.1 eV and larger than that in NiSbS, which is a $3d$ analogue of PtSbS,[27] but remains less than half of BaPtP. Although the crystal structure of BaPtP is different from PtSbS, they have the same space group of $P2_13$. Parameters affecting the magnitude of the band splitting other than spin-orbit coupling, such as the degree of hybridization of wave functions with different parities,[27] are expected to be small in PtSbS.

The electronic structure of PtSbS is featured by unique and complex Fermi surfaces. As shown in Figs. 4(c) and (d), eight bands are involved in Fermi surfaces and there are four sets of surfaces with almost the identical shapes, which are shown next to each other in these figures. Four of them are electron surfaces and the other four are hole ones and they each have small Fermi pockets with an almost identical shape at around the boundary of the Brillouin zone (near the R points for the electron pockets and the M points for the hole pockets). This reflects the degeneracy of electronic states at the zone boundary due to the helical symmetry.[27] In contrast, Fermi surfaces located away from the zone boundary have characteristic shapes varying with each band. Of particular interest is a hole surface with leaf-shaped pockets, shown in the upper panel of Fig. 4(d). Reflecting the 23 point group, they are arranged to swirl when viewed along the 111 direction. In addition, because the leaves are aligned in the $k_x$, $k_y$, and $k_z$ directions, they can overlap well by the parallel shift shown by an arrow in Fig. 4(e), indicative of the strong nesting. Nesting of a Fermi surface gives rise to a charge- and spin-density-wave instabilities and superconducting phases often appear in the vicinity of these instabilities. Such a Fermi-surface instability due to the nesting might play an important role in the emergence of superconductivity in PtSbS.

Interestingly, the Fermi surface topology of the ullmannite-type compounds strongly depends on the materials. For example, the leaf-shaped Fermi surface described above is absent in NiSbS, where the corresponding band is located below the $E_F$ on the Γ-X line. Instead, the same band crosses the $E_F$ in the Γ-R direction in NiSbS, resulting in the complex shape of Fermi surfaces different from those in PtSbS.[27] As expected from Fig. 4(a), the topology of Fermi surface could be greatly changed by a slight difference in chemical composition and structural parameters. This feature in this noncentrosymmteric cubic crystal structure suggests that the ullmannite-type compounds are promising candidates to realize interesting electronic properties, including unconventional superconductivity with various kinds of Cooper pairs and a pressure-induced Lifshitz transition.

In summary, the ullmannite-type PtSbS having a cubic crystal structure without space inversion symmetry is strongly suggested to be a bulk superconductor with $T_c$ = 0.15 K by electrical resistivity and DC magnetization measurements using polycrystalline samples. As far as we are aware, it is the first bulk superconductor in the ullmannite-type compounds. First principles calculations indicate that the spin splitting at around $E_F$ is small in PtSbS despite the electronic states being mainly composed of Pt $5d$ orbitals. Characteristic Fermi surfaces including the hole pockets with strong nesting might be related to the emergence of superconductivity in this compound. We hope that the superconducting properties of PtSbS such as the symmetry of Cooper pairs will be uncovered in the future by experimental and theoretical studies.

## ACKNOWLEDGMENTS

This work was partly carried out under the Visiting Researcher Program of the Institute for Solid State Physics, the University of Tokyo and supported by JSPS KAKENHI (Grant Number: 18H04314, 19H05823, 19K21846, 15H05883, 18H04306).

*E-mail: yokamoto@nuap.nagoya-u.ac.jp